\documentclass[referee,a4paper,fleqn,usenatbib]{mnras}

\usepackage{newtxtext,newtxmath}
\usepackage{media9}

\usepackage[T1]{fontenc}
\usepackage{ae,aecompl}

\usepackage{graphicx}	
\usepackage{amsmath}	
\usepackage{amssymb}	

\newcommand{\Msun}{\mbox{$M_{\odot}$}}
\newcommand{\my}{\mbox{$M_{\odot}$~yr$^{-1}$}}
\newcommand{\kms}{\mbox{km~s$^{\sf -1}$}}
\newcommand{\tgt}{{OH\,231.8+4.2}}

\def\apj{ApJ}

\def\aj{AJ}
\def\aap{A\&A}

\def\pasp{PASP}

\def\wat{H$_{2}$O}

\def\nh3{NH$_{3}$}
\def\kms{km~s$^{-1}$}

\title[Registration of H$_2$O and SiO masers in \tgt]{Registration of H$_2$O and SiO masers in the Calabash Nebula, to confirm the Planetary Nebula paradigm}

\author[R. Dodson et al.]{
R., Dodson,$^{1}$\thanks{E-mail: richard.dodson@icrar.org}
M. Rioja,$^{1,2,3}$ 
V. Bujarrabal,$^{3}$
J. Kim,$^{4}$
\newauthor S.H. Cho,$^{5}$
Y.K. Choi,$^{5}$
Y. Youngjoo$^{5}$
\\
$^{1}$International Centre for Radio Astronomy Research, The University of Western Australia, 35 Stirling Hwy, Western Australia\\
$^{2}$CSIRO Astronomy and Space Science, 26 Dick Perry Avenue, Kensington WA 6151, Australia\\
$^{3}$Observatorio Astron\'omico Nacional (IGN), Alfonso XII, 3 y 5, 28014 Madrid, Spain\\
$^{4}$Shanghai Astronomical Observatory, Chinese Academy of Sciences, Shanghai 200030, China\\
$^{5}$Korea Astronomy and Space Science Institute 776, Daedeokdae-ro, Yuseong-gu, Daejeon, 34055, Republic of Korea\\
}

\date{Accepted XXX. Received YYY; in original form ZZZ}

\pubyear{2017}

\begin{document}
\label{firstpage}
\pagerange{\pageref{firstpage}--\pageref{lastpage}}
\maketitle

\begin{abstract}
%

We report on the astrometric registration of VLBI images of the SiO and \wat\ masers in \tgt, the iconic Proto-Planetary Nebula also known as the Calabash nebula, 
using the KVN and Source/Frequency Phase Referencing. 
This, for the first time, robustly confirms the alignment of the SiO masers, close to the AGB star, driving the bi-lobe structure with the water masers in the out-flow. 
We are able to trace the bulk motions for the \wat\ masers over the last few decades to be 19\,\kms{} and deduce that the age of this expansion stage is 38$\pm$2 years.
The combination of this result with the distance allows a full 3D reconstruction, and confirms that the \wat\ masers lie on and expand along the known large-scale symmetry axis and that the outflow is only a few decades old, so mass loss is almost certainly on-going.
Therefore we conclude that the SiO emission marks the stellar core of the nebular, the \wat\ emission traces the expansion, and that there must be multiple epochs of ejection to drive the macro-scale structure.
\end{abstract}




\section{Introduction}

The largest proportion of stars in the sky, those with initial masses between about 0.5 and 8 \Msun, will reach the AGB phase at the end of their lives. By the end of this phase, the majority of the initial mass will have been ejected, forming a planetary nebulae (PNe). Before then there is a brief proto-PN (pPN) stage. The ejected material is returned to the interstellar medium, enriching its composition (and that of the new generations of stars to be formed from it) with heavy elements.
The copious mass-loss of a AGB star forms a thick circumstellar envelope (CSE) around it. Mass-loss is a basic phenomenon in their evolution, the ejection rate tends to increase with time and, by the end of the AGB phase, it is so strong, as high as 10$^{-3}$ \my, that most of the stellar initial mass is ejected in a relatively short time. Then the AGB phase ends, the stellar core becomes exposed and can be seen. This new star is very compact and shows an increasingly high temperature, rapidly evolving to the blue and white dwarf phases.
At the same time the circumstellar nebula is also evolving very fast: from the nearly spherical CSE in the AGB phase, which is in relatively slow expansion (at, say, 10-15 \kms), to strongly axisymmetric PNe around the dwarf core. A large fraction of the PNe go through a stage with strong axial symmetry and very fast bipolar outflows \citep[etc]{bujarrabal_01, balick_02}. 
This metamorphosis is a very fast and spectacular phenomenon. Within about 1000 yr, the star becomes a blue dwarf able to significantly ionise the nebula, which has already developed a wide bipolar shape. 
It is suggested that the strong axial symmetry typical of post-AGB nebulae is due to the ejection, during the first pPN phases, of very fast and collimated stellar jets. These jets shock the fossil slow-moving AGB envelope, generating a series of axial shocks that cross the massive CSE, inducing high axial velocities in it. The very high amounts of energy and momentum observed in young PNe impose severe limitations on the jet launching mechanism. At present, the only way to explain the origin  of such energetic flows is to assume that a fraction of the ejected CSE is re-accreted by the central star or a companion through a rapidly rotating disk. The jet would then be powered by a magnetocentrifugal launching process, similar to that at work in forming stars. For that to  be efficient the presence of a stellar companion (or at least a massive planet) is necessary, since otherwise the circumstellar material lacks the angular momentum to form an accretion disk.

Maser lines are initially strong in these highly excited AGB circumstellar layers. However as soon as the star+nebula system leaves the AGB phase the mass-loss rate decreases and the 
cool, abundant gaseous materials condense onto the dust. 
Therefore inside the dust layer conditions tend to be too dilute to generate SiO masers
and these are rarely observed. For example, in the survey by \citet{yoon_14}, only 14 out of 164 post-AGB stars were found to have SiO masers, compared with 40 of the 143 AGB stars.
However there is one, paradigmatic, example showing both SiO and H$_2$O masers: the strongly bipolar nebula \tgt, the Calabash Nebula. 

\begin{figure}
\centering
\includegraphics[width=0.7\textwidth,angle=0]{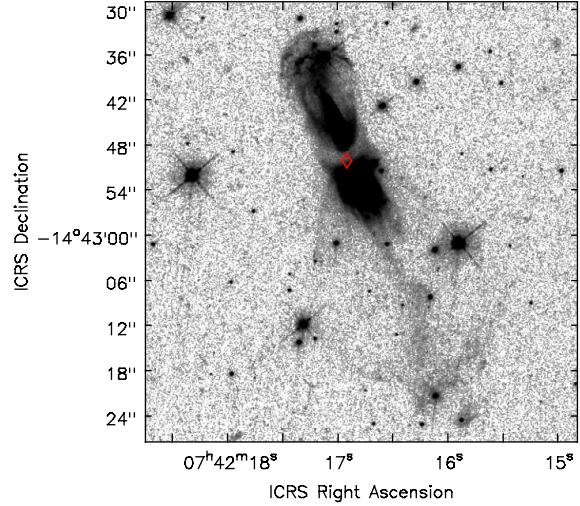}
{\caption
{Hubble image of \tgt\ from the HST Legacy Archive with filter F675, showing the classic bi-polar structure. Marked with a diamond is the region containing the phase referenced positions of the \wat\ and SiO masers, which are clearly at the centre of the expansion of the two lobes.}
  \label{fig:opt}}
\end{figure}

\tgt\ is perhaps the best studied pPN. The optical image, Figure \ref{fig:opt}, reproduces the Hubble image from the Legacy archive, 
and shows the classic bi-polar structure with the axis of symmetry in the plane of the sky of $\sim$21$^o$. 
\tgt\ has a binary central source, which has been identified through optical spectroscopy; a M9-10 III Mira variable (i.e. an AGB star) \citep{cohen_81} and an A0 main sequence companion \citep{sanchez_04}. 
This remarkable bipolar nebula shows all the signs of post-AGB evolution: fast bipolar outflows with velocities $\sim200 - 400$\kms, shock-excited gas and shock-induced chemistry. 
The distance of this source is $\sim$1.54\,kpc  \citep{choi_12b}.
The inclination of the bipolar axis with respect to the plane of the sky, $\sim$36$^o$, is well known, thanks to measurements of phase lags between the variability of the radiation and the light polarization from the two lobes \citep{kastner_92,shure_95}. 

VLBI observations of SiO and H$_2$O masers with the VLBA have yielded a number of important results \citep[for example, see][]{sanchez_02,desmurs_07,choi_12b,leal_12}. Water vapour emission comes from two regions in opposite directions along the nebula axis and presumably the H$_2$O clumps represent the inner nebula, at the base of the bipolar flow. SiO masers occupy smaller regions and lie almost exactly perpendicular to the axis. The Doppler velocities and relative positions of the SiO emission were modelled as a disk orbiting the central star(s) \citep{sanchez_02}. In principle, we are seeing in this object the whole central structure of disk plus outflow that would confirm our ideas on the post-AGB nebular dynamics.
However the astrometric information in these observations is either missing or poor; The SiO observations of \citet{sanchez_02} and \wat\ observations of \citet{leal_12} were self-calibrated, so have no absolute positions. The joint observations of \wat\ and SiO by \citet{desmurs_07} are phase referenced, but the SiO v=2 detection is only `tentative', and no image nor spectrum is provided. 

The import of the relationship of the SiO maser location with respect to the larger scale features is due to the high excitation energies required to invert the SiO transition, which imply that these must mark the location of the central AGB-star \citep{masers_elitzur}.
Therefore, whilst it is logical that the SiO/AGB-star is at the center of the nebulae, between the H$_2$O clusters, definitive astrometry is required to allow conclusions on the disk/outflow association, but efforts over the last decade to provide the registration have been unsuccessful. The main reason for this is that conventional phase referencing at mm-wavelengths is extremely challenging, and that the masers weaken as they expand away from the central star. 
In this paper we present bona-fide astrometric registration of the SiO emission to the positions of the \wat\ masers using Source Frequency Phase Referencing (SFPR), themselves registered to an absolute frame with conventional phase referencing.

\section{Observational Details}

\tgt\ was observed by the Korean VLBI Network (KVN) on 25 Jan, 2017 (N17RD01A) with simultaneous 13mm and 7mm frequencies, each recording 4, dual polarisation, Intermediate Frequency (IF) bands of 16-MHz width. These were spread to cover the \wat\ and v=2,1 J=1$\rightarrow$0 transitions at 22235.044, 42820.57 and 43122.09-MHz respectively and provide the maximum frequency span compatible with the backend (64-MHz at 22-GHz and 382-MHz at 43-GHz), giving accurate delay measurements. 

The KVN offers a truly unique capability; simultaneous frequency phase referencing between bands.
This uses the KVN multi-frequency receiver \citep{kvn_optics} and the SFPR technique \citep{rioja_11a}, which registers the high frequency, mm-wave image against the low frequency image, allowing the measurement of the change of source structure across the frequency bands. Examples demonstrated include core-shifts for AGNs \citep[]{rioja_14,rioja_15,rioja_17a} and spatial relationships between maser transitions \citep[for H$_2$O and SiO]{dodson_14,cho_17}. 

The challenge for astrometric registration of the SiO masers in \tgt\ is that both sets of masers have faded since their first discovery. The reported fluxes in 2009 were 35 and 2.5 Jy for H$_2$O and SiO (J=1$\rightarrow$0, v=2) \citep{sio_h2o_survey}, but the  SiO total flux is falling and now is only 0.5Jy, which is considered too weak for conventional phase referencing. This, in combination with the practical difficulties of phase referencing at 7mm, has hitherto prevented the clear registration of the SiO maser against a calibrator, which could then be compared to the absolute phase referenced H$_2$O masers.
However SFPR allows us to i) use the lower frequency to stabilise the higher frequency data, improving the coherence time and the detectability of the targets and ii) derive the relative astrometric separation of the H$_2$O and SiO masers.

We used 3C84 for prime calibration and J0746-1555 as the reference source, which is 1.5$^o$ from \tgt\ and has 160 mJy at 22GHz and a flat spectrum \citep{petrov_16}.
We used a switching time of 60~seconds for phase referencing, to give matching phase errors from the dynamic and static troposphere at 22-GHz, based on the fomulae in \citet{asaki_07}.
The recording bandwidth was 1Gbps, to ensure a good fringe detection in every scan.
%
The correlator configuration was four bands of 16MHz dual polarisation, for each frequency, and 2048 channels to give 0.1\kms and 0.05\kms at 22 and 43GHz channel width, respectively. 
We note that the phase of the optical cycle (from AAVSO) is approximately 0.5, where the SiO maser emission would be expected to be minimum \citep[see, for example,][]{pardo_04}.

\section{Results and Discussion}

\subsection{Registration and comparison}
\begin{figure}
\centering
\includegraphics[width=0.7\textwidth,angle=-90]{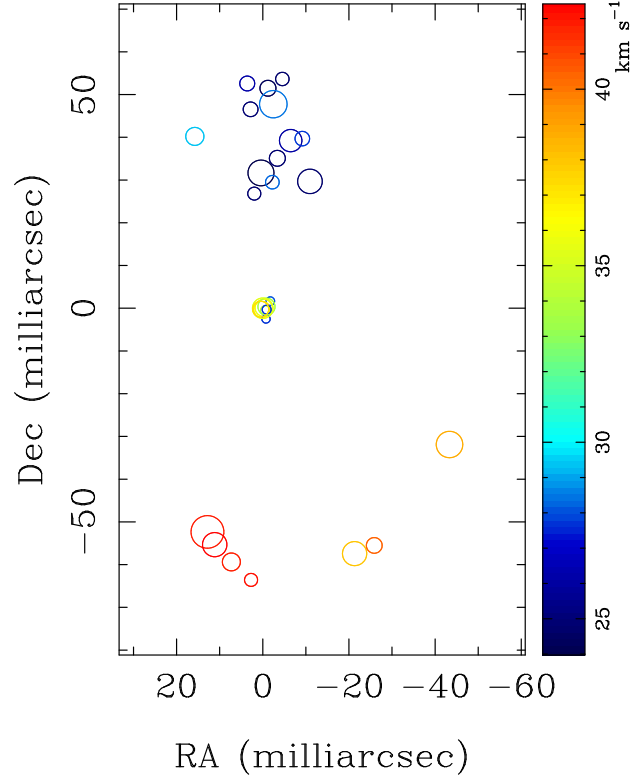}
{\caption
{Phase Referenced spot map of H$_2$O, SiO v=1 and v=2 J=1$\rightarrow$0 masers in OH\,231.8+4.2 on epoch 2017.1, plotted relative to the absolute position 07:42:16.915, -14.42.50.07. The size of each spot is scaled to the flux of the maser and and the colours match the velocity scale on the side bar. The SiO emission is at the centre with the \wat\ emission in clusters to the North and South.}
  \label{fig:d17}}
\end{figure}

Initial instrumental calibration was against 3C84, to remove the clock contributions and align the IFs. The amplitude flux scale was based purely on the measured system temperatures and the assumed dish and receiver system efficiencies. These are reasonably accurate at 22 and 43GHz, and we adopt the canonical 15\% flux scale accuracy\footnote{https://radio.kasi.re.kr/kvn/status\_report\_2017/gain\_calibration.html}. The noise residuals over a 0.2\kms\ channel is typically 38 and 48 mJy/beam at 22 and 43-GHz, respectively.
We measured the residual delays from J0746-1555 at 22GHz, and the phases from a point source model-fit to the strongest channel of the \wat\ maser. These were applied to the whole \wat\ maser dataset. %
These results were also scaled up and applied to the SiO maser datasets, following the method presented in \citet{dodson_14}. Therefore the \wat\ and SiO masers are on a common reference frame. Figure \ref{fig:d17} shows the relative phase referenced positions of the detected emission at epoch 2017.07 in J2000 coordinates.  
The absolute positions of the strongest features are in Table \ref{tab:abs}. 

We lack recent single dish observations from the KVN (the last observation was taken 1.8 years before the VLBI observations), but at that point both v=1 and 2 were detected and the velocities are consistent with our results. We note that the VLBI observations will resolve out at least a fraction of the observed single dish flux, further complicating any comparison. A paper describing the well-sampled KVN single dish monitoring of this source is being prepared \citep{jhkim_17}.
Absolute position errors are those for the calibrator, which is 0.3\,mas.
The relative position error for referencing between 22/43-GHz would be dominated by the absolute position error of the \wat\ maser position and the fractional bandwidth $\Delta\nu/\nu$, that is 0.15\,mas.
However the dominant relative position error between SiO features will be that from the beam size over the SNR for the individual spots. For the strongest feature this is $\sim$0.2mas, but for the median spot flux this is $\sim$0.5mas.
Table \ref{tab:mas} provides the relative positions for all detected maser features.

\begin{table}
\centering
\begin{tabular}{l|c|llrr}
Maser &Velocity&RA&Dec&Fit Error&Astrometric \\
Transition& (\kms)&h:m:s&d:m:s&($\mu$as)&Error ($\mu$as)\\
\hline
H$_2$O & 27.5--29.5 & 07:42:16.91525 & -14:42:50.02167 & 40 & 300\\
SiO v=2 & 33--38 & 07:42:16.915371 &  -14:42:50.06963 & 10 & 200\\
SiO v=1 & 35--38 & 07:42:16.915379 &  -14:42:50.06995 & 10 & 200\\
\end{tabular}
\caption{Absolute positions for the strongest integrated maser features, as observed on 2017/01/25, with the velocity range of the feature and the fitting and the total astrometric errors. The astrometic errors for the \wat\ emission are relative to the ICRF and, for the SiO emission, relative to the \wat\ maser position. \label{tab:abs}}
\end{table}

\subsection{SiO and \wat\ maser alignment}

We find, as expected and suggested tentatively in \citet{desmurs_07}, that the SiO maser emission is placed almost exactly between the two lobes outlined by the \wat\ spots. 
It is difficult to compare our SiO maser positions with those of \citet{sanchez_02, desmurs_07}, because of the lack of astrometry and proper motion in the previous SiO maps. However, an approximate alignment of the various \wat\ observations can be done, see below, and we can compare the SiO positions with present and past observations of \wat\ masers.  

\begin{figure}
\centering
\includegraphics[width=0.5\textwidth,angle=0]{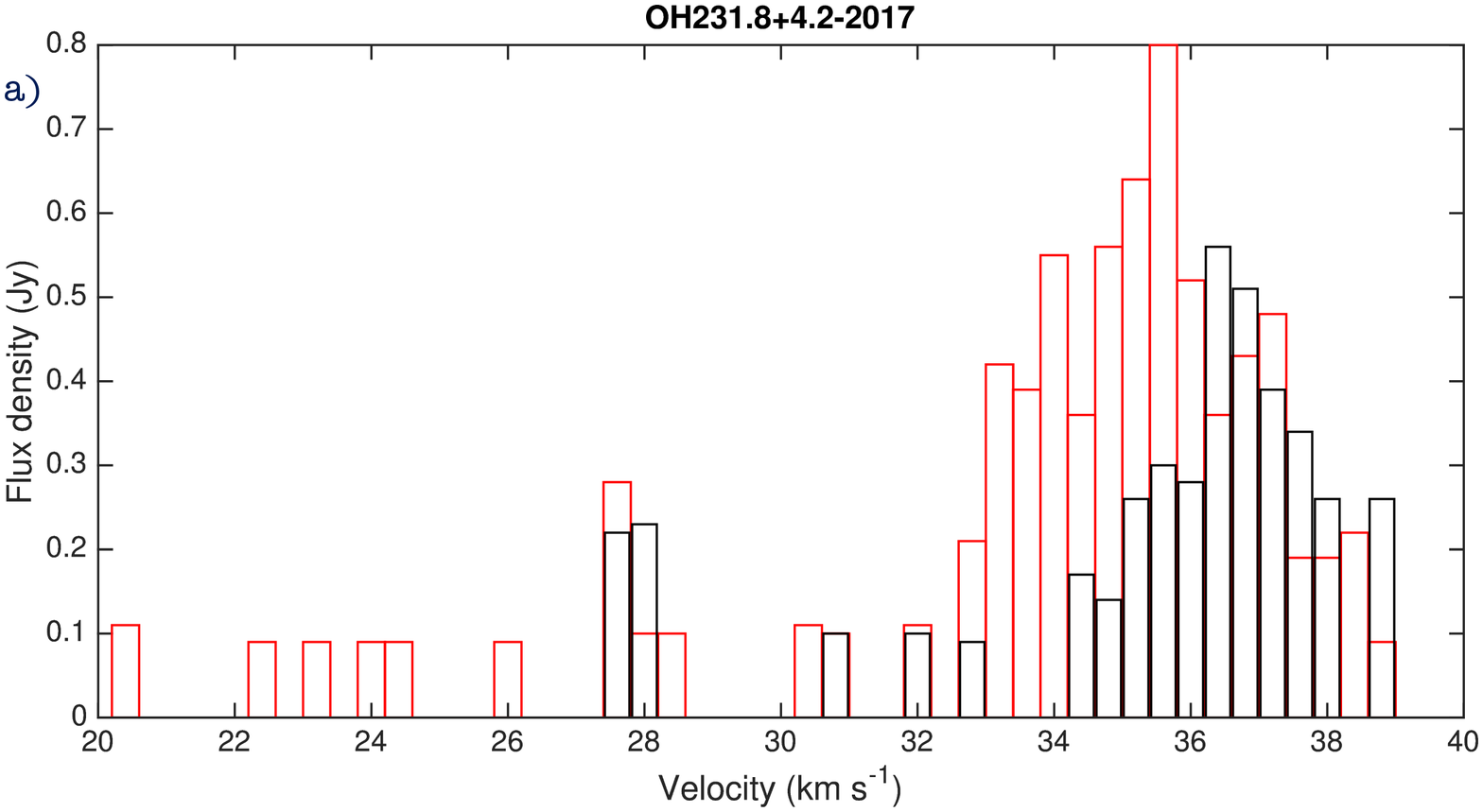}

\includegraphics[width=0.35\textwidth,angle=-90]{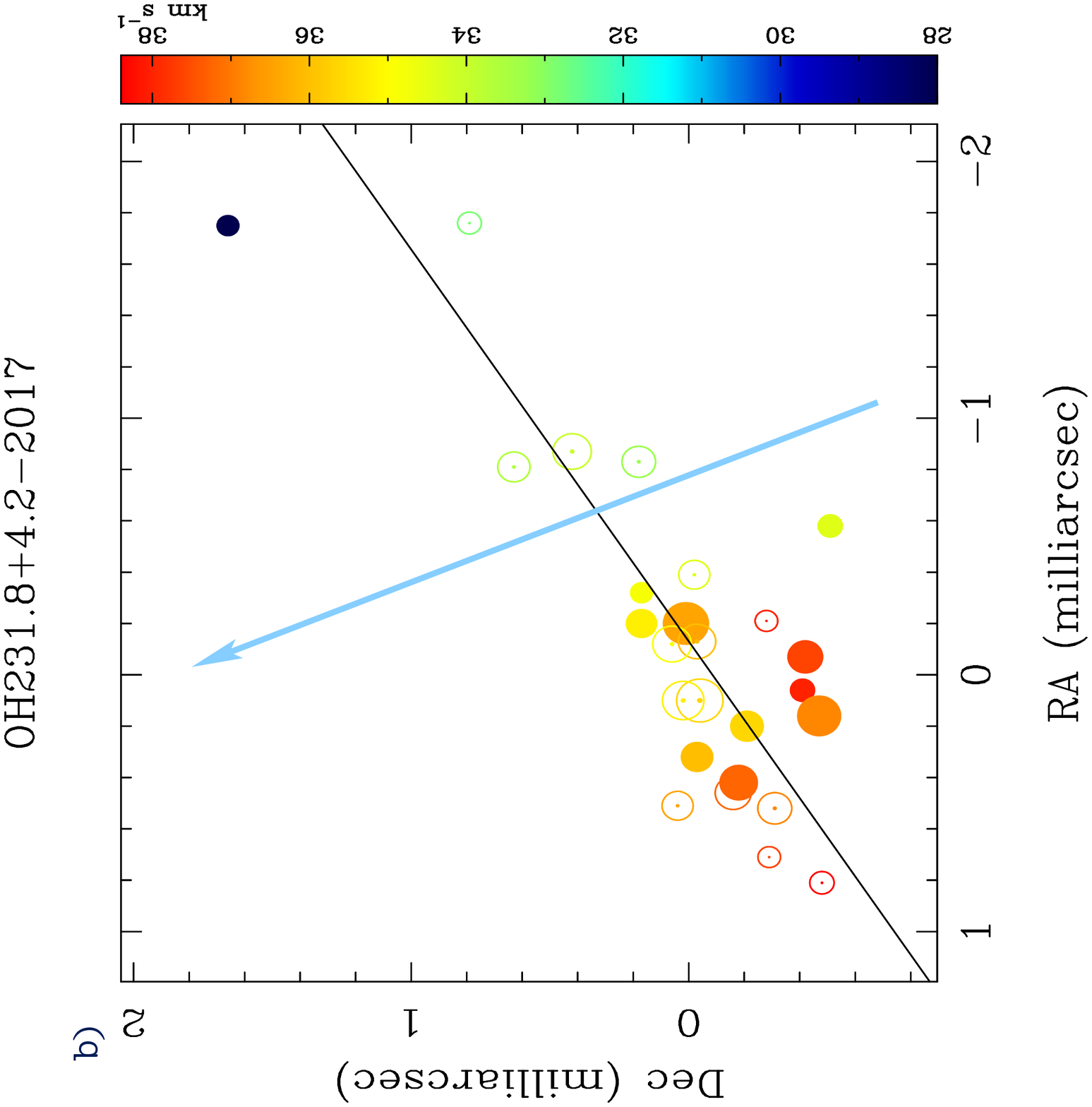}
\includegraphics[width=0.35\textwidth,angle=-90]{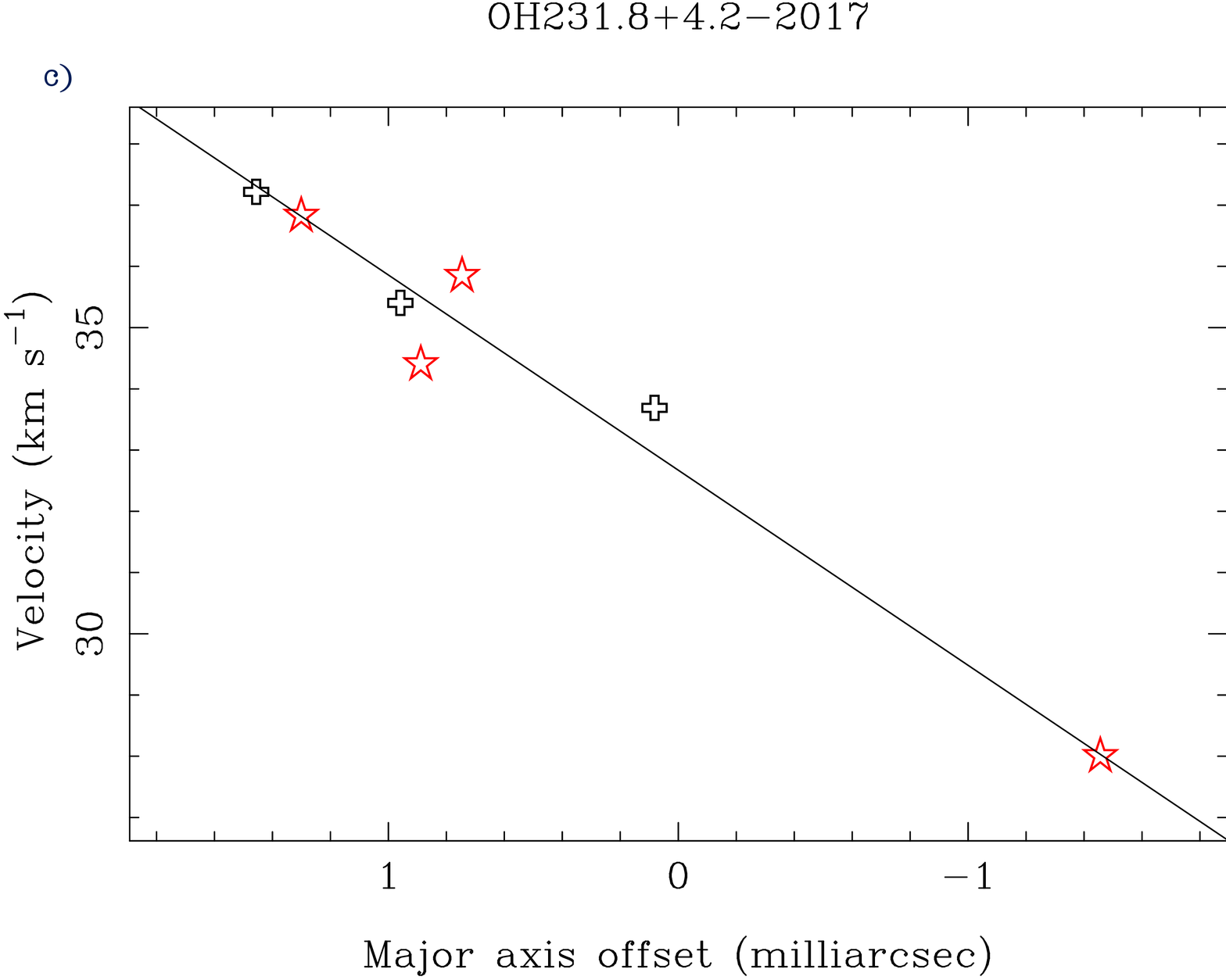}
{\caption {a) The spectrum of the detected spots (formed from the clean components) in SiO v=1 (black) and v=2 (red) masers, clipped at 0.05Jy (1$\sigma$), showing that both transitions have emission around 28 and 36 \kms.
b) The positions of both v=1,2 SiO masers coloured to match the velocity scale on the side bar. All the detected channels that contribute to the spot features in Table \ref{tab:mas} are plotted separately, with v=1 in open circles and v=2 in filled circles. The circle size is proportional to the flux. The light blue line marks the expected axis of symmetry, based on the large scale structure.
c) the velocity-position plots of the SiO maser spots at J=1$\rightarrow$0, v=1 (black cross) and 2 (red star). The enclosed mass, assuming that all emission is detected, would be 0.05 \Msun.}
  \label{fig:vel}} 
\end{figure}

We can see that the SiO masers are placed in the center of the \wat\ distribution, shifted to the North with respect to the \wat\ centroid by $\sim$8\,mas. This result is compatible with the general trend of this nebula to show more extended southern lobes, notably in the wide optical and CO images. The SiO+\wat\ images (Fig. \ref{fig:d17}) are amazingly similar to a reduced version of the optical image (Fig. \ref{fig:opt}), scaled down by a factor $\sim$ 500. Our data also confirm the general structure of the SiO-emitting region, Fig. \ref{fig:vel}: elongated and perpendicular to the nebular axis found at larger scales. We can confirm the detection of an equatorial torus-like structure placed in the very center of this strongly bipolar nebula. If, as usually assumed, the SiO spots are placed in a region tightly surrounding the late-type star, that star is shown to be accurately placed in the center of the nebula.

We are able, because of the quality of the phase referencing and despite the low SiO flux, to detect emission at $\sim$28 and $\sim$36 \kms{} for both the J=1$\rightarrow$0, v=1 and 2 masers. The lines themselves are very broad ($\sim$0.4\kms), and may well be blended because of the KVN resolution, which is 4$\times$3 mas at 43 GHz. However, we did not detect the emission at 40--43 \kms\ found by \citet{sanchez_02} and other authors. The KVN resolution, even at 43 GHz, does not allows a detailed investigation of the structure of the SiO-emitting region. Moreover, the whole emission in our data just occupies about 3 mas (Fig.\ 3), much smaller than the total region detected by \citet{sanchez_02}, which was $\sim$ 8 mas. It is obvious that only a fraction of the torus found by those authors is detected in our data.

The source dynamics are not well probed in our observations.  It could be possible, in principle, to deduce the enclosed mass from the velocity field, assuming Keplerian rotation. However because of the lack of emission at the extreme velocities, the estimate, 0.05 \Msun, is just a lower limit. In fact, this value is very much less than the expected enclosed mass for a M9 or A0 spectral-class star.

\citet{sanchez_02} suggested that the masers lie in a region with a possibly unstable structure and/or kinematics, and they included in their models both rotation and in-fall.  We do see the same general trend in velocity (lowest to the West, highest to the East) as \citet{sanchez_02}. Indeed, it would be possible to place our data within one of their velocity models: that in which both rotation and in-fall are present; their Fig.\ 2, central panel. Under this scenario, we would have detected a 3-mas rim in the South-West part of their structure, with velocities between $\sim$ 30 and 38 \kms. We are aware that this interpretation is very uncertain and we only can conclude that our observations are not incompatible with their model torus.
Deeper higher resolution observations with the VLBA, also using SFPR, will address this.

\subsection{Temporal Evolution}

\begin{figure}
\centering
\includegraphics[width=0.7\textwidth,angle=-90]{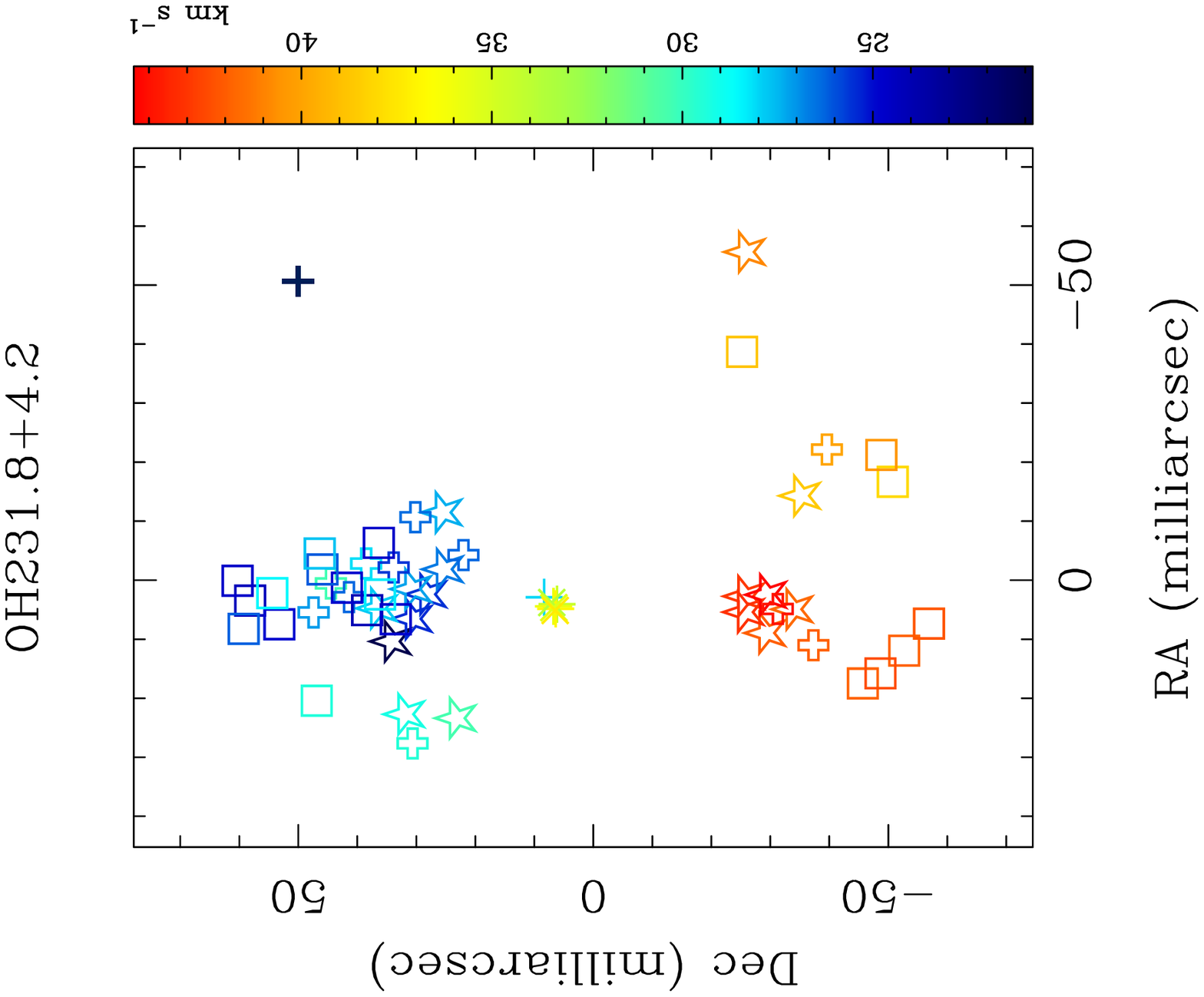}
{\caption
{Non-astrometrical aligned spot map of H$_2$O masers, with data from \citet{desmurs_07} (epoch 2002.9, star) \citet{leal_12} (2009.2, open cross) and this work (2017.1, square), from OH\,231.8+4.2. The different observations, lacking a common reference point, are aligned on the centre of expansion, with an estimated accuracy of 6\,mas. The heavy cross at -50,50 represents the alignment accuracy.
The SiO emission observed in 2017 is also marked (line crosses), registered to the co-observed \wat\ maser emission.}
  \label{fig:all}}
\end{figure}

We have now a number of high quality observations of the \wat\ masers in \tgt, from \citet{desmurs_07,leal_12} and this work. However we do not know, a-priori, the proper motion and therefore we cannot register these observations at different epochs astrometrically.
Nevertheless we can make an approximate alignment by assuming a common centre of the emission, which is plotted in Figure~\ref{fig:all}. We estimate the accuracy of this alignment to be 6\,mas. 
 
We used the change in the separation between the Northern and Southern lobe over time (which does not require absolute astrometry), in Declination, to estimate the age of the nebula. We make a linear fit to both the maximum extent and the median position of the lobes, based on the assumption of a constant expansion velocity. We estimate the birth of the nebula to be 1982$\pm$4 or 1979$\pm$2, respectively. The fit to the median \wat\ maser emission is better, so we take that as the age. This implies a proper motion speed for the water masers of 18\kms\ in Declination, or 19\kms\ along the major axis. This is much lower than the observed Doppler velocities (upto $\sim$400\kms) but consistent with the CO velocities at the centre of the nebula \citep[$\sim$20\kms,][]{alcolea_01,bujarrabal_02}. These references detail how both the extended CO- and H$\alpha$-emitting components show approximately constant velocity gradients, indicating that most nebular material was launched in a short, quasi-explosive event about 800 yr ago. 
The detection (by means of the \wat\ emission) of gas ejected just $\sim$ 38 yr ago shows that strong stellar winds continue to be present after the dominant mass ejection and that, very probably, they are still active to date.

\begin{figure}
\centering
\includegraphics[width=0.7\textwidth,angle=0]{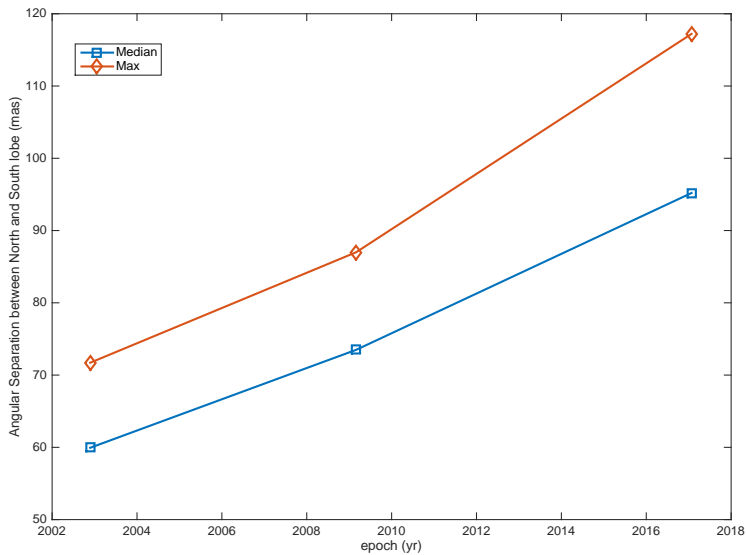}
{\caption
{The internal expansion of H$_2$O masers over 14 years. Shown is the separation between, in red with diamonds, the maximum maser position and, in blue with squares, the median maser position, for the Northern and Southern clusters. 
The expansion of the nebula over the last decade is obvious, and allows us to estimate the age of this structure as being only 38$\pm$2 years.}
  \label{fig:expansion}}
\end{figure}

\begin{figure}
\centering
 \includemedia[
         width=0.9\linewidth,height=0.9\linewidth,
         activate=onclick,
         deactivate=pageclose,
3Dortho=0.003,
3Droll=-1.2,
3Dc2c=-0.9974219799041748 0.07100579142570496 0.01037072017788887,
3Dcoo=1.4210854715202004e-13 1.2256862191861728e-13 -2.511324481702104e-13,
3Droo=166.66667122395828,
         3Dmenu
     ]{\includegraphics[width=0.9\textwidth,angle=0]{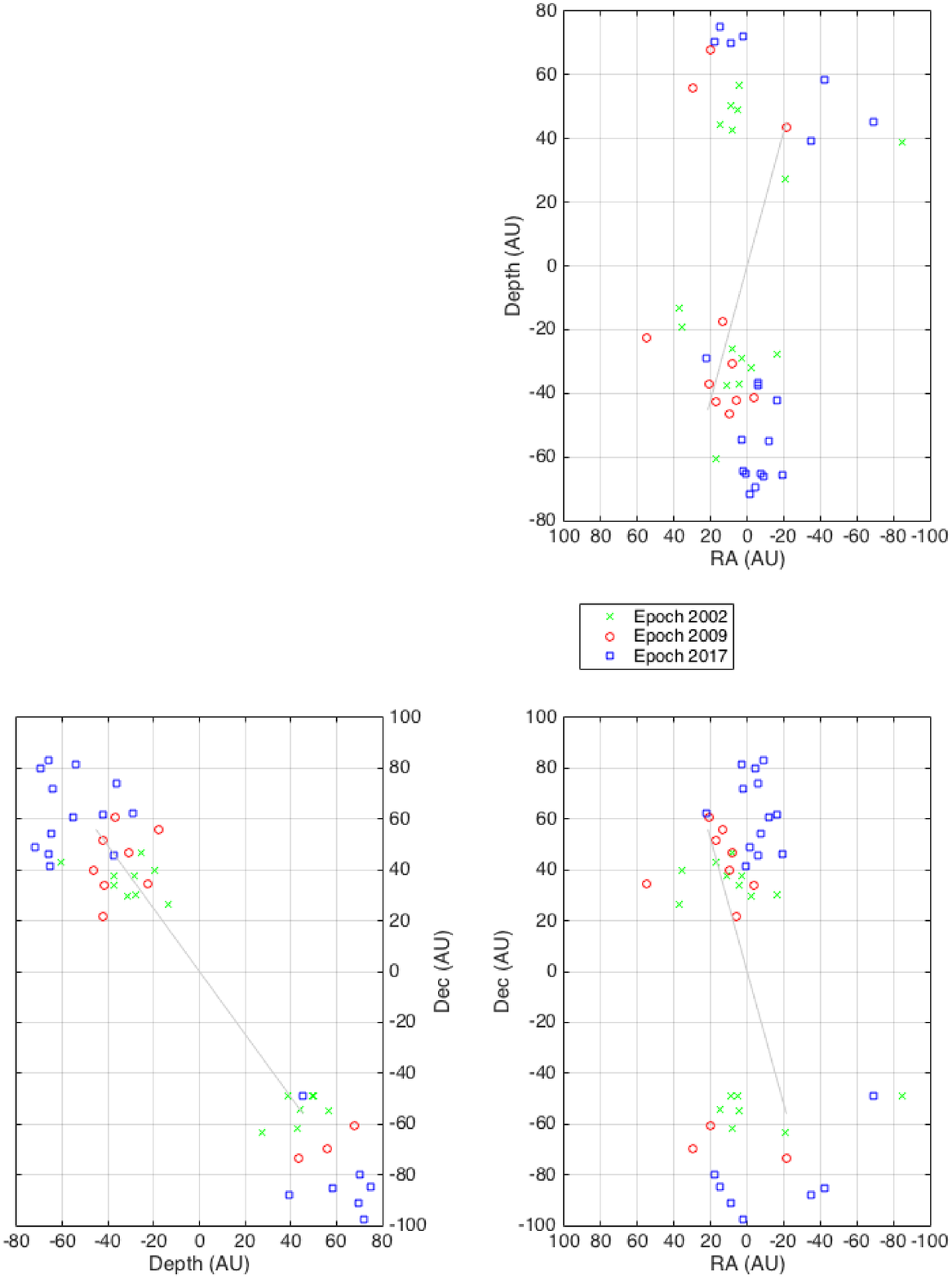}}{Figures/h2o_pos3.u3d}
\caption{The 3D structure of \wat\ masers in \tgt, converted into AU using the distance (for X and Y) and the age of expansion (for Z), for each of the three epochs (2002.9, 2009.2 and 2017.1) plotted in three projections. In light grey is plotted a bar lying along the known 21$^o$ on-sky and 36$^o$ in-sky symmetry axes. 
The emission lies along these axes, thereby allowing us to connect the small scale maser emission to the large scale pPN structure. 
\label{fig:3d}}
\end{figure}

Once we have a zero point (1979) for the expansion of the nebula, and the distance (1.54\,kpc) \citep{choi_12b}, we can convert the velocity relative to the systemic velocity (33\kms) of the emission (spanning 11.4 to -12.2, 10.8 to -7.4 and 9.4 to -9.0\kms\ for epochs 2002.9, 2009.2 and 2017.1, respectively) into depth along the line of sight, and the angular offsets into physical distances in the plane of the sky, for every epoch. We plot the derived positions in Figure \ref{fig:3d}, along with a vector marking the on-sky axis (21$^o$) and deduced inclination angle (36$^o$).
The \wat\ emission shows an excellent alignment with these axes, connecting the small scale maser emission the the large scale structure. 

\section{Conclusion}
We find the peak \wat\ and SiO maser emission, on epoch 2017.07 in J2000 coordinates, to be 07:42:16.91525,-14:42:50.02167 ($\pm$0.3\,mas) and 07:42:16.915379,-14:42:50.06995 ($\pm$0.2\,mas), respectively, with the absolute positional errors given in brackets for the \wat\ maser and the relative positional errors to the \wat\ emission given for the SiO maser.

The crucial result reported in this paper is the registration of SiO and \wat\ masers, which places the SiO only 8mas from the mid point between the two lobes. This, combined with the (previously detected) SiO disk rotating perpendicular around the central star, and the existence of a binary system at the heart of the pPNe, provides compelling evidence that the evolutionary model for bipolar PNe is correct. It also underlines the power of simultaneous multi-frequency observations to provide registration between frequencies in mm-VLBI \citep{dodson_17}.

The SiO emission lies exactly were it would be expected to be, at the centre of the expanding \wat\ emission. Because of the high excitation energies required to invert the SiO transitions, the SiO masers must mark the location 
of the central AGB star. The SiO emission is extended approximately perpendicular to the expansion axis and is probably defining the inner nebula equator. It shows a very different velocity field than that of the matter hosting the H$_2$O maser emission. The SiO-masing component could be in rotation, as proposed by S\'anchez-Contreras et al., but, due to the incomplete detection of the SiO emission, we cannot confirm its dynamics nor estimate the enclosed central mass. 
The KVN observations do not provide sufficient resolution to investigate whether the SiO emission has any peculiar motions associated with in-fall. 

\wat\ observations spanning nearly two decades allows us to estimate the expansion of this outflow in the plane of the sky to be 19\kms\ and therefore the age as being only 38 years. That lifetime is very short, compared with the long time elapsed since most of the large-scale nebula was ejected, demonstrating that the mass loss is on-going. 

Combining the age of the nebula, the distance and the observed maser velocities and positions, allows us to reconstruct the three dimensional structure of the outflow. This shows the inner core of the \tgt\ lies on the axes defined by the optical structures, which map out the long term history of the nebula. Therefore we can conclude that the SiO maser emission, and therefore the central star, is firstly at the centre of the pPN and secondly the \wat\ masers lie on, and are expanding along, the symmetry axis, confirming that these are marking the driving jet that produces these structures. 

\section*{Acknowledgements}
We acknowledge with thanks the variable star observations from the AAVSO International Database, contributed by observers worldwide and used in this research. 
We are grateful to all staff members and students in the KVN who helped to operate the array.  The KVN is a facility operated by the Korea Astronomy and Space Science Institute.
This project has been partially supported by the Spanish MINECO, through grant AYA2016-78994-P.

\begin{footnotesize}

\end{footnotesize}

\begin{table}
\centering
\begin{tabular}{rrr|c}
\hline
RA  & Dec & Velocity & Peak Flux \\ 
 (mas) &  (mas) &  (\kms) &  (Jy) \\ 
\hline
\multicolumn{4}{c}{\wat}\\
\hline 
   0.4 &    31.7 &   23.97  &    1.3 \\ 
  -1.2 &    51.5 &   24.25  &    0.5 \\ 
 -11.0 &    29.7 &   24.72  &    1.1 \\ 
  -3.4 &    35.1 &   24.82  &    0.5 \\ 
   2.8 &    46.5 &   24.89  &    0.4 \\ 
  -4.6 &    53.6 &   24.70  &    0.3 \\ 
   2.0 &    26.8 &   24.81  &    0.3 \\ 
  -6.5 &    39.2 &   26.08  &    0.9 \\ 
   3.6 &    52.6 &   26.16  &    0.4 \\ 
  -9.2 &    39.7 &   27.70  &    0.4 \\ 
  -2.4 &    47.8 &   28.41  &    1.4 \\ 
  15.8 &    40.2 &   29.34  &    0.6 \\ 
  -2.2 &    29.5 &   28.28  &    0.3 \\ 
 -21.3 &   -57.4 &   37.94  &    1.1 \\ 
 -43.3 &   -31.9 &   38.68  &    1.3 \\ 
 -25.9 &   -55.5 &   40.35  &    0.5 \\ 
  12.8 &   -52.3 &   41.83  &    2.0 \\ 
   7.3 &   -59.4 &   41.78  &    0.6 \\ 
   2.7 &   -63.6 &   42.05  &    0.3 \\ 
  11.2 &   -55.3 &   42.42  &    1.1 \\ 
\hline 
\multicolumn{4}{c}{SiO v=2 J=1$\rightarrow$0}\\
\hline 
   -0.9 &    -0.3 &   27.60  &    0.2 \\ 
   -0.8 &    -2.5 &   27.60  &    0.1 \\ 
   -0.9 &     0.4 &   33.69  &    0.6 \\ 
    0.0 &     0.0 &   35.40  &    0.8 \\ 
    0.5 &    -0.2 &   37.22  &    0.5 \\ 
\hline 
\multicolumn{4}{c}{SiO v=1 J=1$\rightarrow$0}\\
\hline 
   -1.8 &     1.7 &   28.00  &    0.1 \\ 
   -0.6 &    -0.5 &   34.40  &    0.2 \\ 
   -0.2 &     0.1 &   35.84  &    0.6 \\ 
    0.2 &    -0.3 &   36.82  &    0.5 \\ 
\end{tabular}
\caption{{Flux weighted Maser spot positions, with reference to absolute position of the peak SiO emission 07:42:16.915, -14:42:50.07.
} \label{tab:mas}}
\end{table}

\bsp	
\label{lastpage}
\end{document}